\documentclass[10pt,twocolumn,letterpaper]{article}

\usepackage{caption}
\usepackage{multirow}
\usepackage{subcaption}

\usepackage{cvpr}
\usepackage{times}
\usepackage{epsfig}
\usepackage{hyperref}       
\usepackage{graphicx}
\usepackage{amsmath}
\usepackage{amssymb}
\usepackage{float}
\usepackage[table]{xcolor}

\usepackage{multirow}

\usepackage{tabularx}
    \newcolumntype{L}{>{\raggedright\arraybackslash}X}
\usepackage[ruled,vlined]{algorithm2e}
\makeatletter
\def\thickhline{%
  \noalign{\ifnum0=`}\fi\hrule \@height \thickarrayrulewidth \futurelet
   \reserved@a\@xthickhline}
\def\@xthickhline{\ifx\reserved@a\thickhline
               \vskip\doublerulesep
               \vskip-\thickarrayrulewidth
             \fi
      \ifnum0=`{\fi}}
\makeatother

\newlength{\thickarrayrulewidth}
\graphicspath{ {./figures/} }

\cvprfinalcopy 

\def\cvprPaperID{****} 

\begin{document}

\title{DeepCOVIDNet: An Interpretable Deep Learning Model for Predictive Surveillance of COVID-19 Using Heterogeneous Features and Their Interactions}

\author{Ankit Ramchandani \\
Texas A\&M University\\
{\tt\small ankitramchandani61@gmail.com}
\and
Chao Fan \\
Texas A\&M University\\
{\tt\small chfan@tamu.edu}
\and
Ali Mostafavi \\
Texas A\&M University\\
{\tt\small amostafavi@civil.tamu.edu}
}

\maketitle
 

\begin{abstract}
In this paper, we propose a deep learning model to forecast the range of increase in COVID-19 infected cases in future days and we present a novel method to compute equidimensional representations of multivariate time series and multivariate spatial time series data. Using this novel method, the proposed model can both take in a large number of heterogeneous features, such as census data, intra-county mobility, inter-county mobility, social distancing data, past growth of infection, among others, and learn complex interactions between these features. Using data collected from various sources, we estimate the range of increase in infected cases seven days into the future for all U.S. counties. In addition, we use the model to identify the most influential features for prediction of the growth of infection. We also analyze pairs of features and estimate the amount of observed second-order interaction between them. Experiments show that the proposed model obtains satisfactory predictive performance and fairly interpretable feature analysis results; hence, the proposed model could complement the standard epidemiological models for national-level surveillance of pandemics, such as COVID-19. The results and findings obtained from the deep learning model could potentially inform policymakers and researchers in devising effective mitigation and response strategies. To fast-track further development and experimentation, the code used to implement the proposed model has been made fully open source.
\end{abstract}

\section{Introduction}
COVID-19 has had an unprecedented social and economic impact worldwide. With more than 13 million infected cases and more than half a million deaths as of mid-July, the pandemic is still accelerating globally without showing any signs of nearing an end. In dealing with COVID-19 and future pandemics, it is imperative to design reliable intervention strategies and to implement effective mitigation efforts. The design of such strategies hinges on effective surveillance of the spatiotemporal evolution of disease. Hence, a reliable and relatively interpretable method of forecasting the spread of the virus could significantly improve the predictive surveillance capability and help in designing policies for disease containment.

Facing this need, prior research has proposed and tested various statistical, epidemiological and machine learning-based forecasting models for COVID-19 \cite{tomar2020prediction} \cite{perc2020forecasting} \cite{anastassopoulou2020data} \cite{huang2020multiple} \cite{chimmula2020time} \cite{rustam2020covid} \cite{punn2020covid} based on features such as number of existing infections, deaths and recoveries. Other forecasting models proposed for other epidemics \cite{siriyasatien2018dengue} \cite{Balcan21484} \cite{wang2019defsi} rely on features like human mobility and within-season and between-season observations. Although these models show potential for predicting the initial outbreak and growth trajectories, their capability in capturing various temporally dynamic and spatially variant features affecting disease spread is limited \cite{jewell2020predictive}. For example, mathematical models such as susceptible-infectious-recovery (SIR) models only account for a small subset of relevant features identified to be responsible for the spread of the virus, as shown below.

Existing studies \cite{holmdahl2020wrong} have shown that the spread of disease is dependent on many factors; thus, reliable forecasting models must capture all major factors that might influence the spread of infection. Specifically, a multitude of factors including, but not limited to, human mobility \cite{kraemer2020effect}, social distancing guidelines \cite{jia2020modeling} \cite{jarvis2020quantifying}, weather \cite{qiu2020impacts}, population density \cite{rocklov2020high}, and demographics \cite{dowd2020demographic} affect the spread of COVID-19. In other fields of study where prediction also depends upon a large number of features, researchers often find that identifying interactions between input features is essential to yielding satisfactory results \cite{song2019autoint} \cite{gao2014machine}. Although, to the best of our knowledge, no literature has yet been published that identifies specific feature interactions for the spread of COVID-19, we hypothesize that it is quite likely that many features upon which the spread is dependent interact in complex ways. Hence, an exhaustive examination of the relevant features and their possible interactions is essential for an effective prediction model of disease spread.

Advances in deep learning can enable contemporary models to use a large number of features and to account for possible interactions. Several deep learning models \cite{guo2017deepfm} \cite{cheng2016wide} \cite{wang2017deep} used for online ad click predictions are particularly known for using multiple heterogeneous features as input and learning interactions among them; however, since many features used for epidemic forecasting are in the form of multivariate time series and multivariate spatial time series, an effective method to compute representations of such features that accounts for both the temporal and spatial structure of data is essential before using them as input to the aforementioned models. Perhaps, due to the challenges associated with building such representations for heterogeneous input features, existing deep learning models for epidemic forecasting rely upon more conventional recurrent architectures that do not explicitly account for feature interactions and consider only few input features \cite{chimmula2020time} \cite{tomar2020prediction} \cite{tian2020forecasting}.

To complement existing epidemiological models, we propose a deep learning model based on the high-level framework of DeepFM \cite{guo2017deepfm} that takes in multiple features, accounts for interactions between them and forecasts growth in the number of infected cases in all U.S. counties. For effective processing of the many input features, the model includes a novel method to compute equidimensional representations (also called embeddings) of heterogeneous features such as multivariate time series, multivariate spatial time series, and multidimensional time-independent variables. Furthermore, we perform feature importance evaluations to identify the most influential features for predicting the growth of infected cases. Also, since the proposed model accounts for possible interactions among input features, we perform an analysis to estimate the relative amount of second-order interaction between pairs of input features. The results show that the model obtains satisfactory performance. In addition, the highly interpretable feature importance results can also help policy makers develop control strategies in response to the rapidly evolving pandemic situation. To fast-track future research and experimentation with new features or models, we have also made our code fully open source.

\section{Related Work}
\label{sec:rel-work}
\subsection{Standard disease-spread models}
To estimate the growth of infected populations, epidemiologists and mathematical modelers have developed multiple statistical and mathematical models to simulate the spread of disease in terms of susceptible, infected, recovered, and deceased populations. These models include the susceptible-infectious-recovery (SIR) model and its derived models, such as the susceptible-exposed-infectious-recovery (SEIR) model \cite{McCluskey2010} \cite{fan2020network}. Through contact activities of people, these standard models attempt to capture the community spread of disease \cite{Liu2018a}. While these models provide useful insights regarding the initial outbreak and growth trajectories, they have limitations in terms of the number of influencing factors and complex relationships captured. For example, existing models can include only a limited number of features (primarily infection rate) to forecast the spread of infection; however, research has shown that disease spread is related to a large number of factors (such as socio-demographic factors, mobility, population density, and visits to points of interest), which possibly interact in complex ways. Data-driven models can be adopted to capture various dynamic features and their interactions to complement the standard disease-spread models. 

Existing studies also reveal that population flow drives the spatiotemporal distribution of COVID-19 cases \cite{Jia2020}, and travel ban was projected to be successful in slowing the epidemic spread. This has been demonstrated in the context of COVID-19 in China \cite{Chinazzi2020}. To complement existing mathematical disease spread models, the global epidemic and mobility model \cite{Balcan21484} was developed to incorporate the movement of people (which may hasten transmission of the disease across different areas) and predict the spread of disease. This model focuses mainly on cross- and within-community transmission of the disease through the analysis of global human mobility; however, as shown in recent literature, the spread of the disease is affected not only by human contact, but is also related to multiple additional factors, such as population density, human shopping activities, and directives of local government \cite{jia2020modeling} \cite{jarvis2020quantifying} \cite{kraemer2020effect} \cite{engle2020staying}. Existing models have a limited capability to capture the effects of these factors; hence, developing a model that can take various relevant factors into account, predict the spread of disease, and attribute importance to each factor could be informative for pandemic surveillance and associated policy making.

\subsection{Deep learning models}

\begin{table*}
    \caption{Short description of input feature groups. Refer appendix \ref{sup:ftr} for full details}
    \begin{tabular}{p{4.5cm} p{8cm} p{4.5cm}}\\
    \thickhline
    \textbf{Feature Group Name} & \textbf{Brief Description}  & \textbf{Type}   \\
    \hline
    Census features &   \small{Population distribution with age/sex, race, poverty/employment status, etc}     &    Constant \\
    Vulnerability features &   \small{Population density and other features to determine vulnerability of a county to COVID-19 (refer appendix \ref{sup:ftr})}    &  Constant \\
    Past rise in infected cases &    \small{Daily rise in cumulative cases in past days}    &	Time-Dependent \\
    Reproduction Number &   \small{Estimate of number of other people infected by one person}    &	Time-Dependent \\
    Venables distance \cite{louail2014mobile} & \small{A measure of the concentration of population activities in a county (refer appendix \ref{sup:ftr})}  &	Time-Dependent \\
    Social distancing metrics &    \small{Amount of adherence to social distancing guidelines}    &	Time-Dependent \\
    Visitation patterns &    \small{Intra-county mobility to certain types of places (grocery stores, amusement parks, etc)}    &	Time-Dependent \\
    Cross-county mobility and infections & \small{Incoming traffic from one county to another and cumulative infected cases in source county}    &	Cross-County Time-Dependent \\
    \thickhline
    \end{tabular}
    \label{tab:in-features}
\end{table*}

Several researchers have identified that predictive models based on deep learning could be effective in aiding decision making in response to COVID-19 \cite{hussain2020ai} \cite{pham2020artificial} \cite{boccaletti2020modeling}. Huang et al. \cite{huang2020multiple} proposed a convolutional neural network to forecast the spread of the virus using six input features related to the number of confirmed, recovered, and deceased people. Their model predicts the spread one day ahead using data from last five days. Although their work shows that deep learning can be effective in forecasting the pandemic, the model cannot be directly deployed in practice because it predicts only one day of future scenario and considers only a limited number features affecting the spread of disease.

Chimmula et al. \cite{chimmula2020time} proposed a long short-term memory (LSTM)  model \cite{hochreiter1997long} to forecast the spread of the virus based on past number of confirmed, deceased, and recovered cases. Similarly, Tian et al. \cite{tian2020forecasting} also used an LSTM model to forecast the spread based on similar features but normalized by population. The authors also compared their results with a Hidden Markov Model and a Hierarchial Bayes Model, but concluded that LSTM exhibits the best performance of the three, demonstrating the potential impact deep learning can have on pandemic forecasting. 

Tian et al. \cite{tian2020covid} also proposed a custom model to forecast cumulative confirmed cases and deaths by combining the LSTM \cite{hochreiter1997long} and GRU \cite{chung2014empirical} cells. Their model used past numbers of confirmed cases and deaths as input and five other time-independent features. They reported their model performance only in terms of relative error, so it is difficult to judge the effectiveness of their model in absolute terms. Further, the authors did not clarify why some features, such as violent crime rate, were used in the model and how these features could contribute to prediction of infections and deaths by COVID-19.

Further, an important and perhaps similar work is the DeepCOVID model \cite{deepcovid}. The associated researchers have developed a deep learning-based forecasting model using ``syndromic, clinical, demographic, mobility, and point-of-care data'' to forecast mortality and hospitalization in the United States. However, only a small amount of information (such as accuracy, model architecture, and feature importance) about this work has yet been made public.

To the best of our knowledge, the proposed DeepCOVIDNet model is among the first significant deep learning models for COVID-19 forecasting to be completely open-source. Moreover, a notable distinction of our work is that we perform feature analysis to understand which features are important in forecasting the growth of the virus, an analysis not performed in prior studies. Further, we also account for possible interactions among our many input features and identify pairs of features with relatively higher amount of second-order interaction between them, which again has not been done by other models.

\section{Input Feature Groups}
\label{sec:in-features}

To have a comprehensive understanding of the situation and characteristics of counties, it is important to examine several features for each county. To this end, we extensively surveyed the literature and identified certain ``influencing factors'' that might affect the spread of infection. We then identified specific feature groups that corresponded with the set of influencing factors identified earlier. A feature group is simply a set of similar features grouped together to facilitate further processing. A brief description of all feature groups is presented in table \ref{tab:in-features}. In this section, we describe the process of feature collection, organization, and inclusion in the proposed model in detail.

\subsection{Feature Collection}
\label{ssec:feature-rationale}
Our extensive analysis of the literature helped us identify four factors that may influence the spread of COVID-19 both spatially and temporally. These ``influencing factors'' include population attributes (such as population density, age/race-based population distribution, etc.), population activities (such as visits to certain types of points of interest, adherence to social distancing guidelines, etc.), mobility (movement from more infected places to less infected ones), and disease spread attributes (such as reproduction number, growth of infections in past, etc.). The feature groups used in correspondence with each influencing factor are shown in table \ref{tab:themes}.  A more complete discussion of all feature groups is provided in the appendix \ref{sup:ftr}. In the following, we provide evidence to show the importance of each influencing factor in predicting the growth of the virus.

Features related to population attributes, such as population size and density, are important in predicting the growth of infection as shown by research studies discussed below. Rockl\"{o}v and Sj\"{o}din \cite{rocklov2020high} demonstrated that the spread of the virus in a geography is directly proportional to its population density. Dyer \cite{dyer2020covid} and Kirby  \cite{kirby2020evidence} showed that the impact of the pandemic has been disproportionately higher in black and other minority communities. Dowd et al. \cite{dowd2020demographic} illustrated how population age distribution and inter-generational contacts (possibly captured by household type and size) affect the impact of the virus in a community. Together, these studies provide a strong rationale for using features that capture population attributes of counties by showing that these unchangeable population characteristics contribute significantly in understanding the spread/impact of the virus. As shown in table \ref{tab:themes}, to account for population attributes, we use census features, population density, and some other engineered features built by the Surgo Foundation to assess vulnerability of each county due to COVID-19 (refer appendix \ref{sup:ftr}).

Multiple research studies have found that population activities are important in predicting the growth of future cases.  Benzell et al. \cite{benzell2020rationing} identified risks associated with visits to venues at which people would be placed in proximity and showed that particularly high risk was associated with visits to restaurants, grocery stores, fast food establishments, cafes, and gyms. Also, Lai et al. \cite{lai2020covid} found that there is high risk of small disease outbreaks within residential facilities for elderly, indicating that examining the number of visits to such facilities could be helpful in forecasting the growth of infection. Moreover, since research has shown that COVID-19 is communicated via airborne particles, especially indoors, visits to hospitals, especially by healthcare workers, could be risky \cite{bahl2020airborne} \cite{morawska2020can}, and therefore are important to consider when determining the growth of infection.  Further, Gao et al. \cite{gao2020mobile} showed that adherence to social distancing orders helps to reduce the spread of infection. Cohen and Kupferschmidt \cite{cohen2020countries} and Sen-Crowe et al. \cite{sen2020social} also acknowledged the importance of social distancing in slowing the spread of the virus. Further, B{\'e}land et al. \cite{beland2020short} demonstrated that workers in certain occupations in which they are more likely to work in proximity are more affected by the virus, indicating that the percentage of people working full- or part-time could influence infection spread. These studies, together, show the importance of capturing population activities as input features. As shown in table \ref{tab:themes}, we use points of interest (POI) visitation patterns, Venables distance \cite{louail2014mobile}, and social distancing metrics to capture these activities. As explained in appendix \ref{sup:ftr}, visitation patterns capture the number of visits to important types of POIs, Venables distance captures the agglomeration of population activities in a county, and social distancing metrics show adherence to stay-at-home orders. Together, these three feature groups show the extent to which people are likely to come in close contact and potentially facilitate the spread of infection.

Further, the dynamics of urban mobility are also important in predicting the spatial spread of the virus. Kraemer et al. \cite{kraemer2020effect} showed that human mobility played a major role in explaining the initial spatial distribution of infections in China. They showed that more than 50\% of initial cases identified outside Wuhan could be traced back to travel from Wuhan. Sirkeci et al. \cite{sirkeci2020coronavirus} also confirmed that human mobility from more infected places to less infected places is a significant factor in predicting the spatial spread of the virus. Other studies \cite{gatto2020spread} \cite{Jia2020} also concurred regarding the role played by mobility in the spread of the virus. Therefore, we use data of incoming traffic from other counties to capture mobility in the proposed model as shown in table \ref{tab:themes}. We further augment the mobility information by including the number of infections in source counties because travel statistics alone are not sufficient to inform the model about counties from which travel could be more dangerous due to the high prevalence of infection.

Finally, disease spread attributes, such as the growth of cases in the past few days and reproduction number, are also important factors in determining the spread of virus in the future. As mentioned in section \ref{sec:rel-work}, several existing models \cite{chimmula2020time} \cite{tian2020forecasting} \cite{huang2020multiple} obtain satisfactory results in forecasting future cases by using features similar to the growth of cases in past days. Further, epidemiology-based studies \cite{fan2020effects} \cite{liu2020reproductive} \cite{dietz1993estimation} have reported that estimating the reproduction number can be essential to understanding the behavior of virus spread, thereby making it an important input feature to consider. Based on these studies, we include both the growth of cases in past days and reproduction number (estimated by the method proposed by Fan et al \cite{fan2020effects}) as input features to the model.

\begin{table}
    \caption{The feature groups used in correspondence with each influencing factor}
    \begin{tabular}{p{2.5cm} p{4.75cm}}\\
    \thickhline
    \textbf{Influencing Factor} & \textbf{Feature Groups Contained}  \\
    \hline
    \multirow{2}{7em}{Population attributes} &   a) Census features \\ 
    & b) Vulnerability features \\[0.25cm]
    
    \multirow{3}{7em}{Population activities} & a) Visitation patterns \\
    & b) Venable's distance \\
    & c) Social distancing metrics \\[0.25cm]
    
    Mobility & a) Cross-county mobility \\[0.25cm]
    
    \multirow{2}{9em}{Disease Spread Attributes} & a) Reproduction number \\ 
    & b) Past rise in infected cases \\
    
    \thickhline
    \end{tabular}
    \label{tab:themes}
\end{table}

\subsection{Feature Organization}
As discussed above, we combine similar features into a feature group to better process and organize data. We define three types of feature groups based on their spatial and temporal characteristics as described below: constant, time-dependent, and cross-county time-dependent feature groups. The exact feature groups used in the model are shown in table \ref{tab:in-features} and are fully explained in the appendix \ref{sup:ftr}. 

\textbf{Constant Feature Groups:} Constant feature groups contain features that do not vary significantly over the analysis period. For example, the set of census features for a county (such as population size and population density) are considered constant features because their values do not change significantly within a few months. 

\textbf{Time-Dependent Feature Groups:} This group contains features whose values change over time. For example, the number of people who visit grocery stores on a particular day in a given county is a time-dependent feature since its value changes depending on the particular day. A time-dependent feature group is essentially a multivariate time series. 

\begin{figure*}
    \centering
    \includegraphics[scale=0.53]{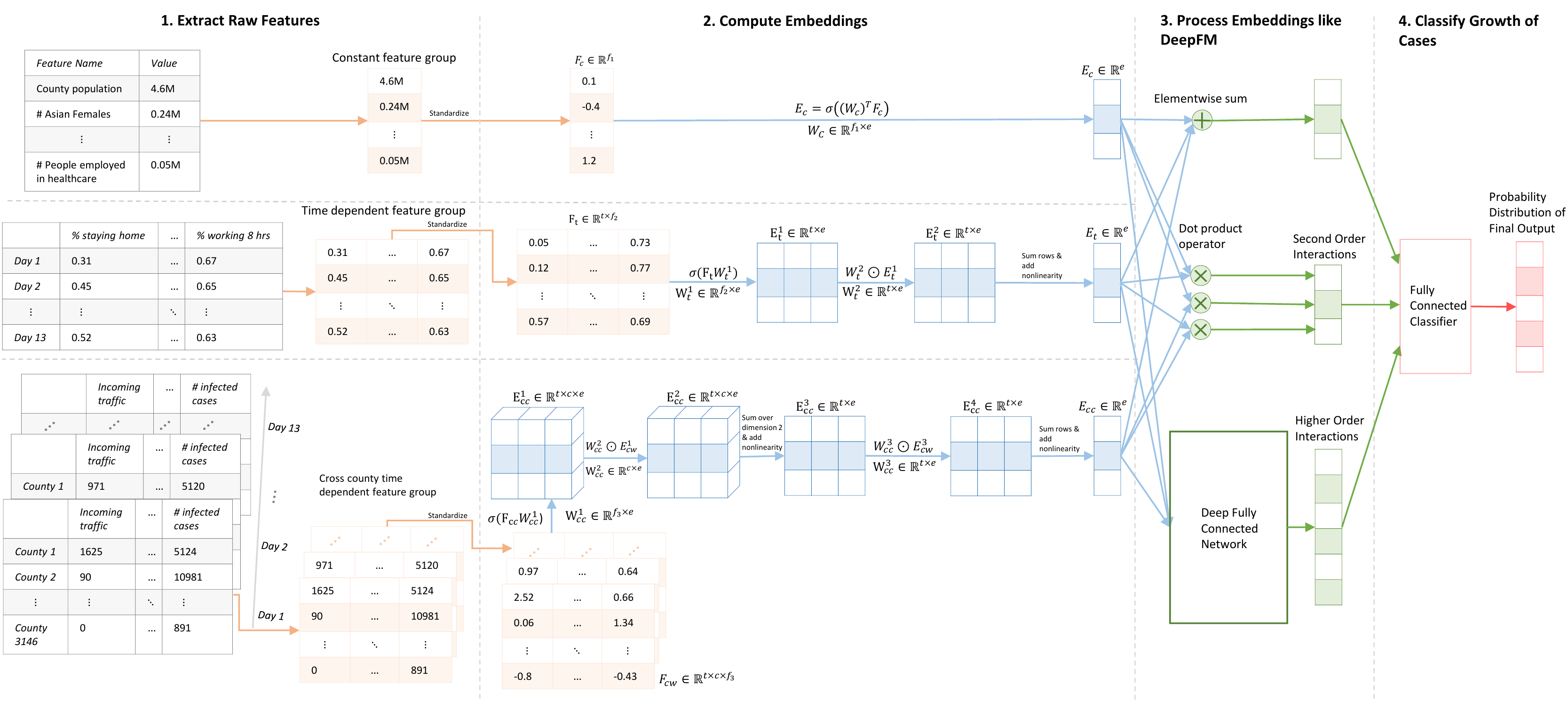}
    \caption{A schematic representation of the entire pipeline of the proposed model which takes as input raw features and outputs a probability distribution to predict the range of the rise in infected cases in a given county on a given date in the future. $f_1, f_2, f_3$ correspond to number of features, $t$ refers to the number of time steps (experimentally set to 13), $c$ refers to the number of counties (set to 3146), $e$ refers to the dimension of each embedding, and $\sigma$ refers to any non-linearity. More explanation of tensor shapes is presented in \ref{ssec:emb}.}
    \label{fig:model}
\end{figure*}

\textbf{Cross-County Time-Dependent Feature Groups:} In cross-county time-dependent feature groups, features capture a time-dependent interaction between two counties. For example, the number of people traveling from one county to another on a particular day is a cross county time dependent feature. A cross-county time-dependent feature group is essentially a multivariate spatial time series.

\section{DeepCOVIDNet}
\label{sec:model}
We have designed a novel deep learning model, DeepCOVIDNet, to estimate the range of increase in the number of infected cases on a particular day given multiple constant, time-dependent, and cross-county time-dependent feature groups as input. The ``projection interval'' of the model is 7 days, which means the model can forecast the range of increase in cases for 7 days into the future.

The model comprises of two modules: the embedding module and the DeepFM module. The embedding module takes as input various heterogeneous feature groups and outputs an equidimensional embedding corresponding to each feature group. These embeddings are then passed to the DeepFM module which computes second- and higher-order interactions between them. Finally, we use a shallow, fully connected network which takes as input the computed interactions and the sum of feature embeddings (to improve gradient flow) and outputs $n$ probabilities corresponding to $n$ binary classification tasks, as explained in section \ref{sec:impl}.  We transform the $n$ probabilities to get the probability distribution of the current rise in cases to lie within $n+1$ ordered ranges. We expect the rise in cases to lie within the range with the greatest probability. Figure \ref{fig:model} shows a schematic representation of the process described above.

\subsection{Embedding Module}
\label{ssec:emb}
The novel embedding module used in this study produces embeddings of the same dimension for all feature groups using a generalizable process, which is described in detail below for each feature group type. The fundamental idea behind extracting embeddings from various feature groups is to utilize all available structure in the data. For example, for time-dependent features, we would like the model to be able to treat the same features on different dates more similarly than different features on different dates. We accomplish this goal by sharing parameters, which is a successful technique used to design novel models and arguably is at the heart of the success of both recurrent and convolutional neural networks.

\subsubsection{Constant Feature Embeddings} 
Since constant feature groups do not have a time dimension, the shape of a group of constant features is simply $[n]$, where $n$ is the number of features in the group. As shown in figure \ref{fig:model}, the embeddings of constant features are calculated simply by using a fully connected layer that converts the input tensor of shape $[n]$ to a tensor of shape $[e]$ where $e$ is the embedding dimension.

\subsubsection{Time-Dependent Feature Embeddings} 
\label{p:time-dep-emb} 
Since the time dimension of time-dependent feature groups needs to be taken into account, the shape of a time-dependent feature group is $[t, n]$, where $t$ represents the time steps, and $n$ represents the number of features for each time step. From the $n$ input features, our method computes a holistic feature score by performing a weighted summation of the $n$ original features for every time step. This holistic score can be thought of as a new time-dependent feature, engineered by the model as per its needs. Next, a weighted summation is conducted over all time steps to learn the influence of different time steps on the final output. As shown in equation \ref{eq:time-dep}, when the embedding size is $e$, the model has a chance to engineer $e$ new time-dependent features and understand how each new feature behaves over time.

In a formal formulation, let $F_t \in \mathbb{R}^{t \times n}$ be a time-dependent feature matrix of shape $[t, n]$. Let $({F_t})_{ij}$ represent the $j^{th}$ feature at the $i^{th}$ time step. Let the output embedding be $E_t \in \mathbb{R}^e$ and $(E_t)_k$ represent the $k^{th}$ element in the embedding. Then, we calculate $(E_t)_k$ in the following way:

\begin{align}
    \label{eq:time-dep}
    (E_t)_k = \sigma(\sum_{i \in [1 \dots t]} W^T_{ik} \cdot \sigma[\sum_{j \in [1 \dots n]} W^F_{jk} \cdot (F_t)_{ij}])
\end{align}
where $\sigma$ is any activation function and $W^F \in \mathbb{R}^{n \times e}$ and $W^T \in \mathbb{R}^{t \times e}$ are learnable weights in the network.

In part 2 of figure \ref{fig:model}, we show exactly how equation \ref{eq:time-dep} is implemented in a vectorizable method.

\subsubsection{Cross-County Time-Dependent Feature Embeddings} 

Since each feature value of a cross-county time-dependent feature group is associated with all counties and all time steps, the shape of a cross-county time-dependent feature group is $[t, c, n]$, where $t$ represents the total time steps, $c$ represents the total number of counties in the U.S. and $n$ represents the number of features.

As shown in figure \ref{fig:model}, let $F_{cc} \in \mathbb{R}^{t \times c \times n}$ represent the raw features so that $(F_{cc})_{ijk}$ is the $k$\textsuperscript{th} feature of the $j$\textsuperscript{th} county at the $i$\textsuperscript{th} time step. Further, if we let $E_{cc} \in \mathbb{R}^e$ represent the final embedding of this feature group and $(E_{cc})_p$ represent the $p^{th}$ element of $E_{cc}$, then we compute $(E_{cc})_p$ using an extension of equation \ref{eq:time-dep} in the following way:

\begin{multline}
    \label{eq:c-time-dep}
    (E_{cc})_p = \\
    \sigma(\sum_{i \in [1 \dots t]} W^T_{ip} \cdot \sigma(\sum_{j \in [1 \dots c]} W^C_{jp} \cdot \sigma[\sum_{k \in [1 \dots n]} W^F_{kp} \cdot F_{ijk}]))
\end{multline}
where $\sigma$ is any activation function, $W^F \in \mathbb{R}^{n \times e}$, $W^C \in \mathbb{R}^{c \times e}$ and $W^T \in \mathbb{R}^{t \times e}$ are learnable weights in the network.

Note that equation \ref{eq:c-time-dep} is an extension of equation \ref{eq:time-dep} over the county dimension.  As in the previous case, the innermost summation computes a holistic feature score for every county and time step using the same weights. Next, an overall score for each time step is computed by a weighted sum of all holistic feature scores for all counties. Finally, a weighted combination of all time steps is computed.

\subsection{DeepFM Module}
We hypothesize that there exist several interactions between different feature groups. An interaction between two features exists when their values together convey some information that cannot be extracted by considering their values individually. For example, there could be an interaction between the percentage of population staying indoors in a county and the total number of infected cases in the county. A higher incidence of new cases is expected if few people remain indoors and there are already a high number of infected cases. However, the model may not be able to predict the number of new cases as effectively if only one of the two feature values is given. Hence, it is important for the model to identify and learn many such interactions that could exist among different features. In this section, we provide an overview of how the model computes second- and higher-order interactions among input features using the DeepFM \cite{guo2017deepfm} framework (with slight modifications).

The embeddings of all features obtained from the embeddings module described above serve as input to the DeepFM model. Note that all raw features groups are of different sizes, but their embeddings have the same dimensions and are easy to further process by the DeepFM module, as shown in figure \ref{fig:model}.

In the model, the dot products between a pair of embeddings represent second-order interactions between the corresponding two feature groups. To identify higher-order interactions, we concatenate all embeddings and process them through a self-normalizing neural network \cite{klambauer2017self}. The network comprises of a sequence of dense layers with the same output dimensions, followed by the SELU (Scaled Exponential Linear Units) non-linearity and the alpha dropout layer \cite{klambauer2017self}. We treat the number of dense layers and the dimension of their output as hyperparameters in the model, which are tuned by using Bayesian optimization as explained in section \ref{sec:impl}.

\section{Implementation Details}
\label{sec:impl}
\textbf{Loss function.} The proposed model predicts a range between which the rise in the number of infected cases is expected to lie. In this section, we describe how the model is trained with this output. Let $C \in \mathbb{Z}^n$ represent a list containing boundaries of ranges used for prediction in the model. Therefore, if the model predicts class $i \in [0, n]$, the rise in cases is expected to be within the interval $[C_i, C_{i + 1})$ under the assumptions that $C_0 = 0$ and $C_{n+1} = \infty$. Naturally, $C_i \le C_{i + 1}$ for all $i$.

Since $C$ is sorted and ordered, we can treat the output of the model as an ordinal variable and perform ordinal regression using the method described by Frank et al. \cite{frank2001simple}. According to the method, $n$ binary classifiers need to be trained such that classifier $i \in [1, n]$ outputs the probability $P(x > c_i)$, where $x$ is the rise in number of cases and $c_i$ is a constant. After $n$ probabilities are obtained from these $n$ classifiers, we can easily find $P(c_j \leq x < c_{j + 1})$ for all $j \in [0, n]$. Finally, the rise in cases is expected to lie in the interval $[c_k, c_{k + 1})$ for some $k \in [0, n]$ so that $ \forall_i\ P(c_k \leq x < c_{k + 1}) \geq P(c_i \leq x < c_{i + 1})$. Frank et al. \cite{frank2001simple} suggest that the model should be trained by using only binary cross entropy loss on the $n$ binary classifiers. However, this could seem non-ideal because the final goal is to predict the interval $[c_j, c_{j + 1})$ rather than to predict $P(x > c_i)$ for all $i$. Therefore, for the proposed model, a multi-class cross-entropy loss on $P(c_i \leq x < c_{i + 1})$ for all $i$ is added in addition to the binary cross-entropy loss on all binary classifiers. The conducted experiments show that the results are slightly better when this new loss term is added.

\begin{figure}
\centering
\includegraphics[scale=0.35]{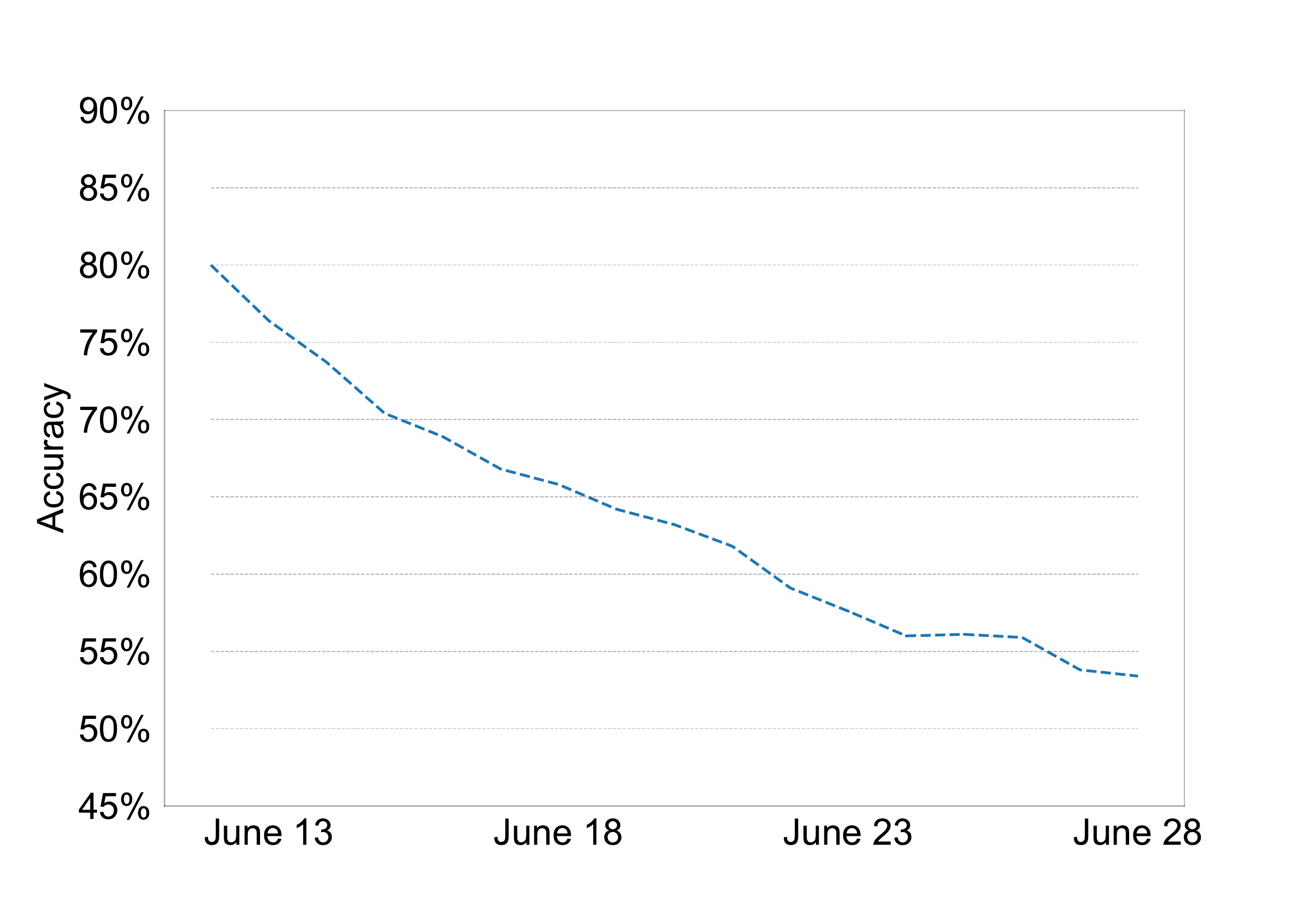}
\caption{The plot of accuracy vs time on the testing set - the results show that the accuracy keeps decreases as the model tries to project further in the future.}
\label{fig:acc}
\end{figure}

\textbf{Class ranges for outputs.} As described above, the proposed method predicts a range within which the increase in the number of cases would lie. It is important to discuss how range boundaries (i.e., list $C$) are actually chosen. Basically, we start by finding the rise in the number of cases in every county from April 5 through June 28 during the same projection interval of the model. We then find the $33$\textsuperscript{rd}, $67$\textsuperscript{th} and the $90$\textsuperscript{th} percentile of the rise in cases during those dates. In raw numbers, this turns out to be 1, 13, and 93. So the output classes correspond to the following ranges: 0 to 1 (negligible increase), 2 to 13 (moderately low increase), 13 to 93 (moderately high increase), and 93 and above (significantly high increase). These numbers denote the rise in cases in every county during one week. Further, it also should be noted that due to the distribution of the labels, a naive model which predicts the same class always would at most get 33\% accuracy. This should also help contextualize our results. We achieved about 64\% accuracy on the testing data (June 12 through June 28), which is about two times better than a naive model. In light of this and the feature analysis results, we can be confident that the model has learned useful information. It is also important to note that these class boundaries can easily be changed to make the model predictions finer or coarser as per different use cases.

\textbf{Train and test splits.} The total data used was from April 5 through June 28. 68\% of this data was used for the training set, 12\% for validation and 20\% for the testing set. Equivalently, data from April 5 through June 1 was used for training, June 2 through June 11 for validation, and June 12 through June 28 for testing. Although data since January 21 was available, we used data only from April 5 onward because of the lack of widespread testing availability before then. Since the model was trained on labels derived from the results of COVID-19 testing data, it was essential to use data that was relatively accurate and did not underestimate the number of infections due to lack of testing \cite{jewell2020predictive}.

\begin{figure*}
\centering  
\begin{subfigure}{.5\textwidth}
    \includegraphics[width=1\linewidth]{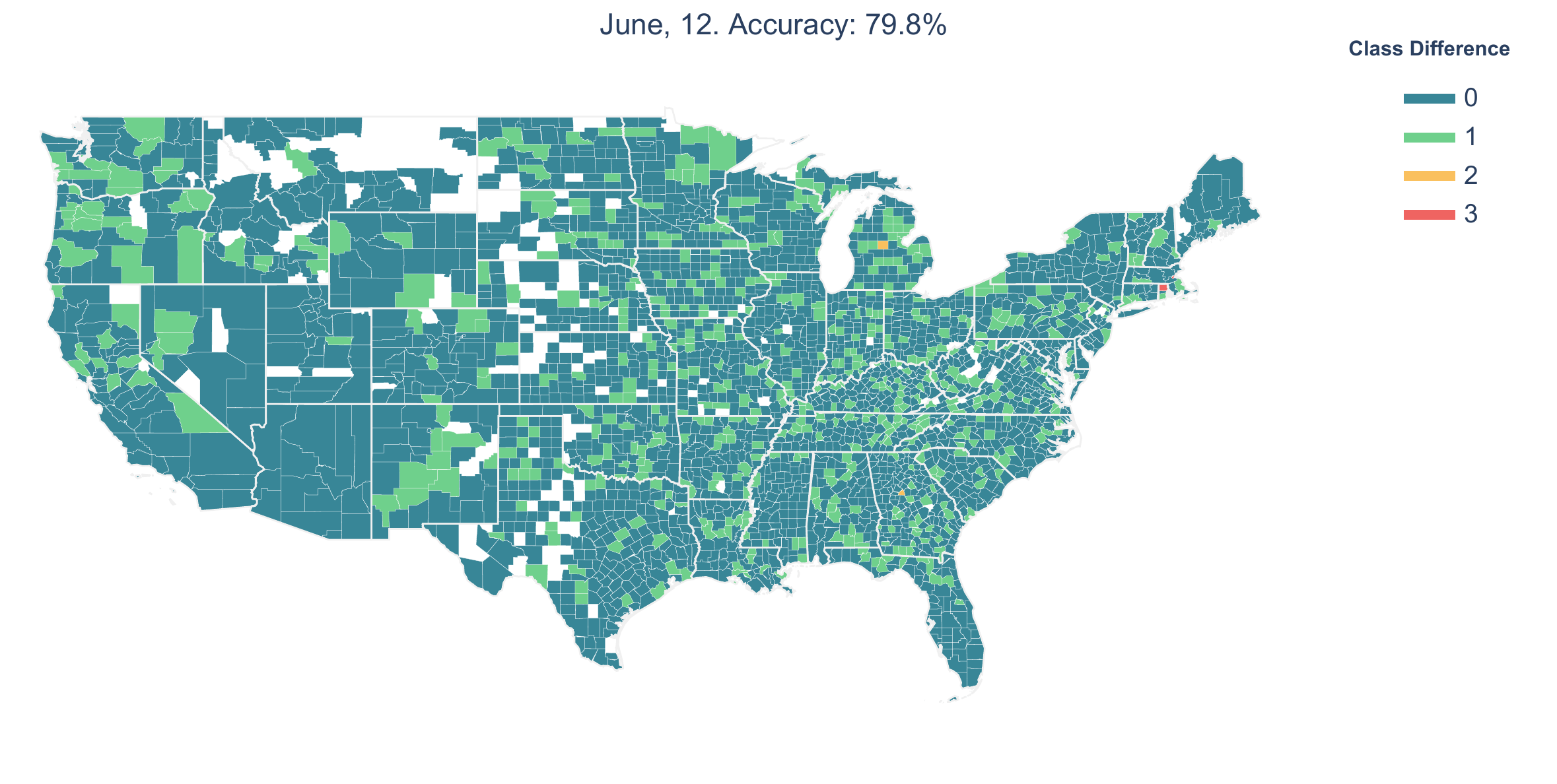}
\end{subfigure}%
\begin{subfigure}{.5\textwidth}
    \includegraphics[width=1\linewidth]{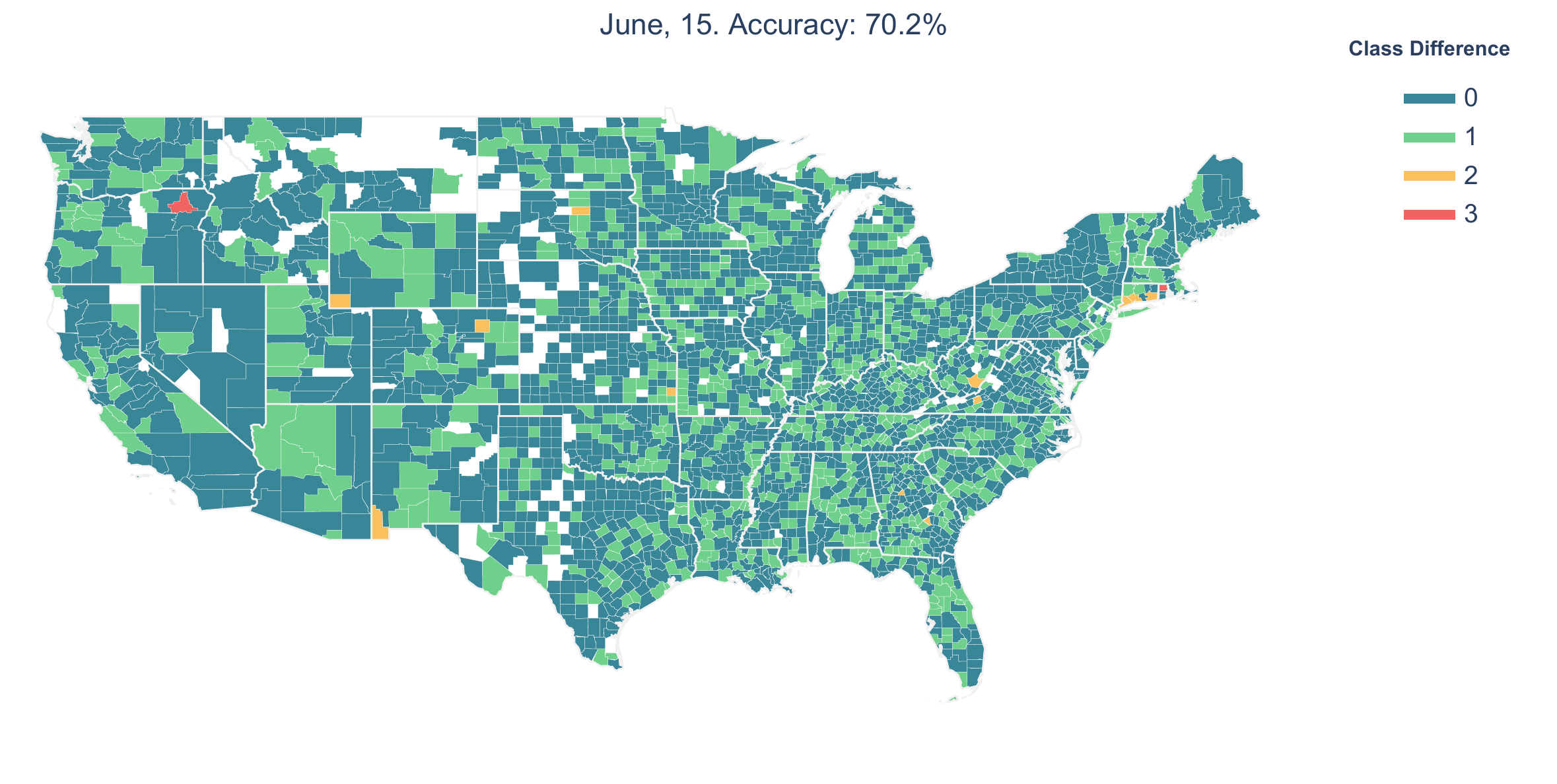}
\end{subfigure}%
\hfill
\begin{subfigure}{.5\textwidth}
    \includegraphics[width=1\linewidth]{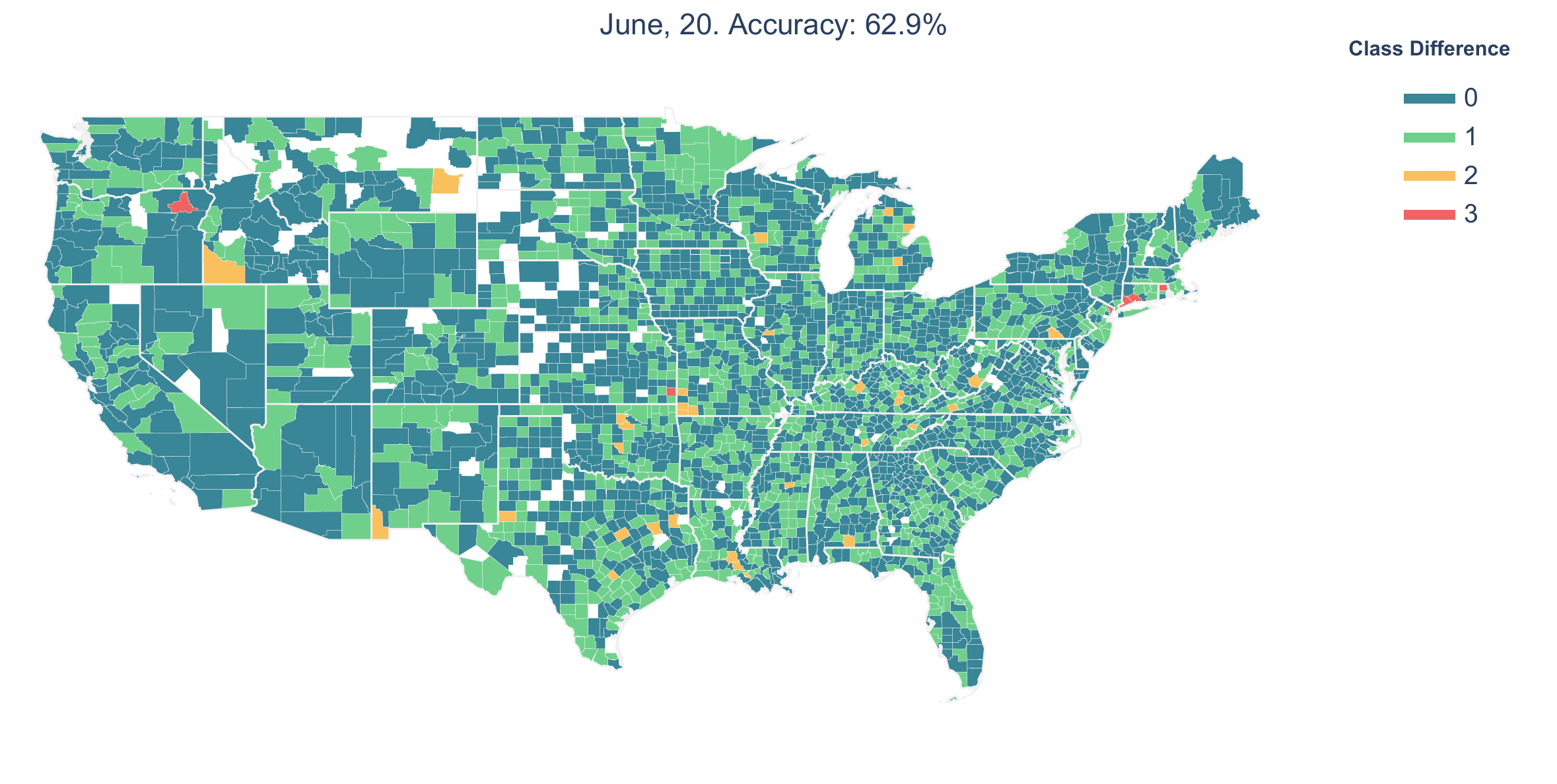}
\end{subfigure}%
\begin{subfigure}{.5\textwidth}
    \includegraphics[width=1\linewidth]{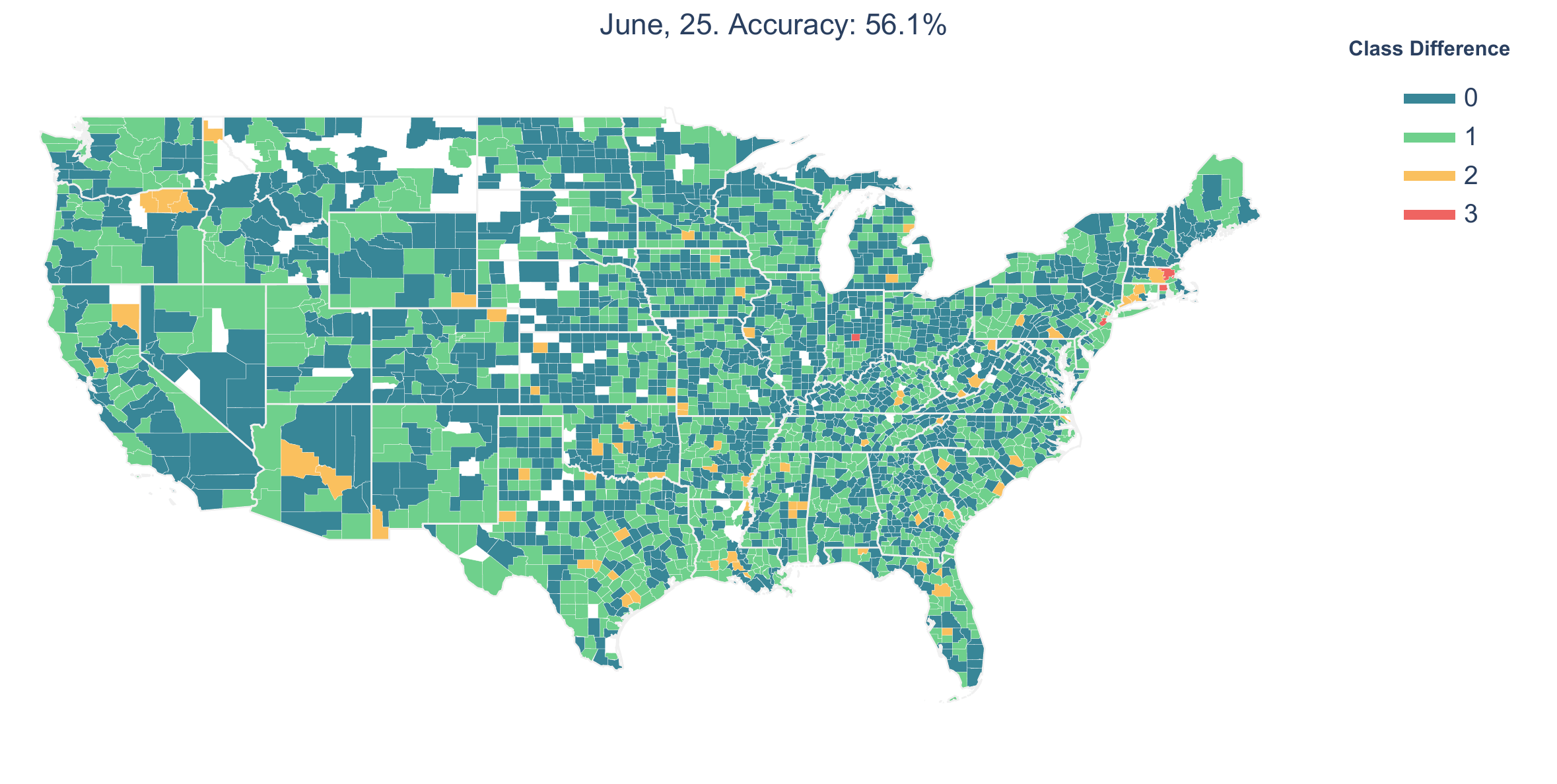}
\end{subfigure}%
\caption{Results showing the absolute difference in predicted class and the actual class for all counties (since classes are ordered). Therefore, the value of 0 means the model's prediction was correct, 1 means it was one class away from ground truth, and so on. The accuracy of the model on the given date is shown above each subfigure. (The few counties not shown in these plots either have no infected cases or did not have sufficient data.)}
\label{fig:spatial}
\end{figure*}

\textbf{Amount of Past Data Used.} As explained in section \ref{sec:in-features}, the values of many input feature groups used in the model change with time (time-dependent or cross-county time-dependent feature groups). For all such features, we used 13 days of past data. We experimentally found that increasing the number of past days to 21 and 28 had insignificant effect on the model accuracy. Therefore, for reasons related to computational efficiency, we chose to use 13 days of past data.

So, to be clear, when predicting the range of increase in infected cases on, for example, June 21, we use input data from June 1 to June 13.  Features from June 14 through 20 are not given as input because that represents the interval for which the model is predicting the rise in cases.

\textbf{Hyperparameter Tuning.} We used Bayesian optimization for 30 iterations with expected improvement as the acquisition function to choose hyperparameters for the model \cite{snoek2012practical}.

\textbf{Training Method.} As shown in figure \ref{fig:acc}, the model's testing results become less accurate as it is tested on dates further in time from the dates on which it was trained. Since the dates of the training, validation, and testing set are in ascending order, we hypothesized that training on the validation set as well should improve testing performance, since more recent dates will be used in training. In light of this belief, the model is trained in two steps.

First, we keep the original training and validation sets intact and use Bayesian optimization as described above to perform hyperparameter tuning and choose the model with the best performance on the validation set. We note the best hyperparameters and the number of epochs needed to train the best model. Second, we train a fresh model on both the validation and training sets with the hyperparameters chosen in the first step and for the same number of epochs. The model obtained in the second step is the final model. Note that the first step has two purposes: (1) choosing best hyperparameters, and (2) helping decide a stopping point for the second step to avoid/reduce overfitting.

\begin{figure*}
\centering
\begin{subfigure}{.33\textwidth}
    \centering
    \includegraphics[scale=0.18]{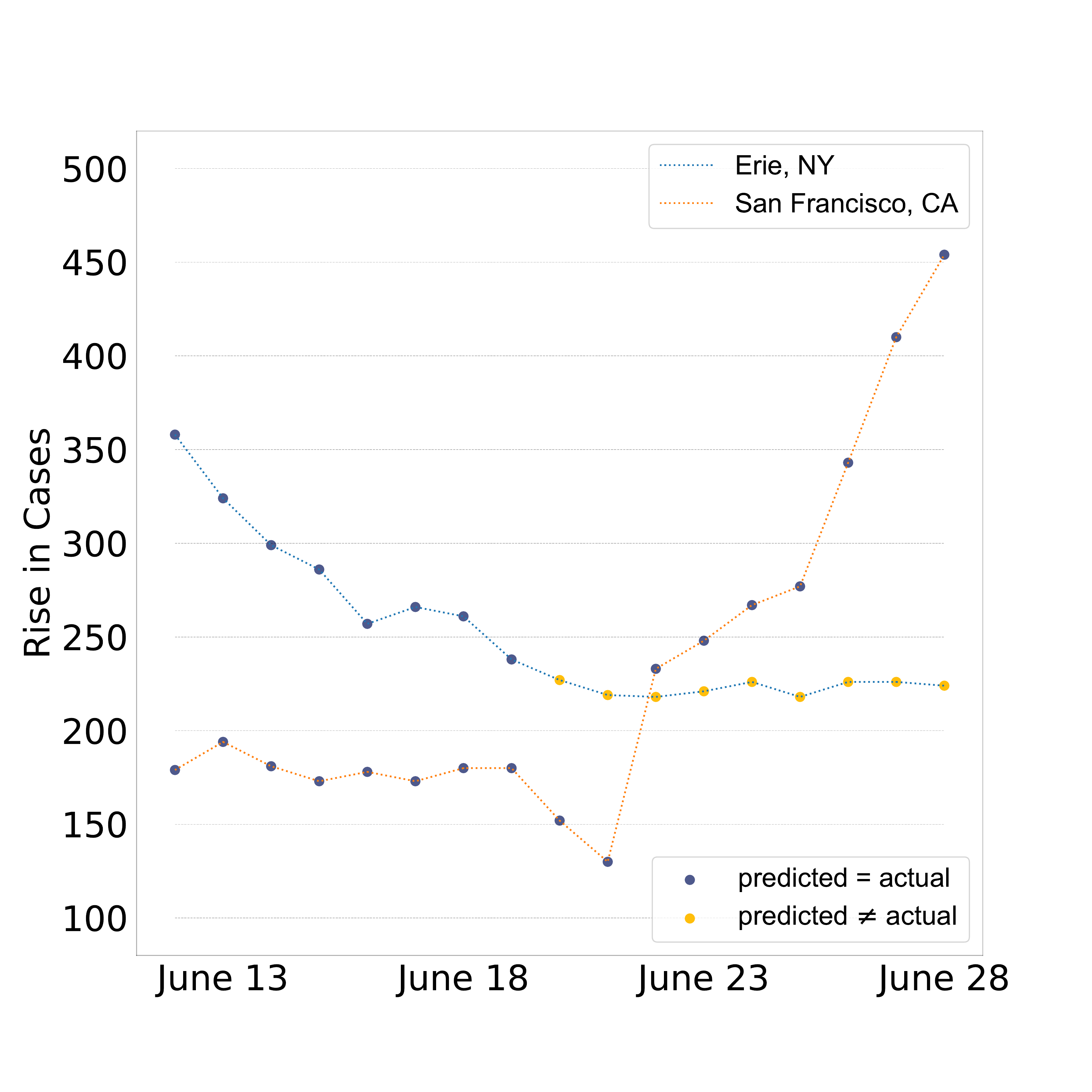}
\end{subfigure}
\begin{subfigure}{.33\textwidth}
    \centering
    \includegraphics[scale=0.18]{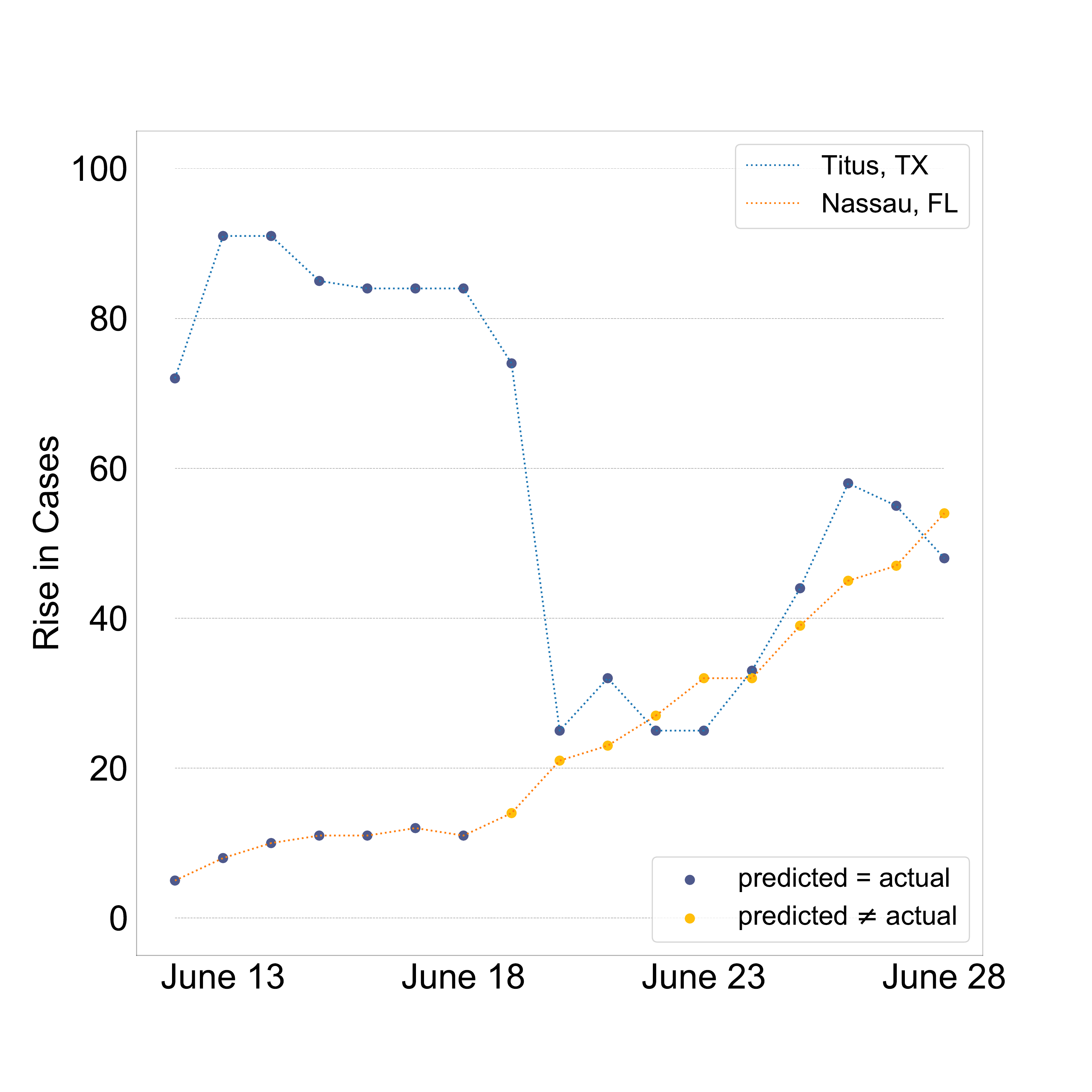}
\end{subfigure}
\begin{subfigure}{.33\textwidth}
    \centering
    \includegraphics[scale=0.18]{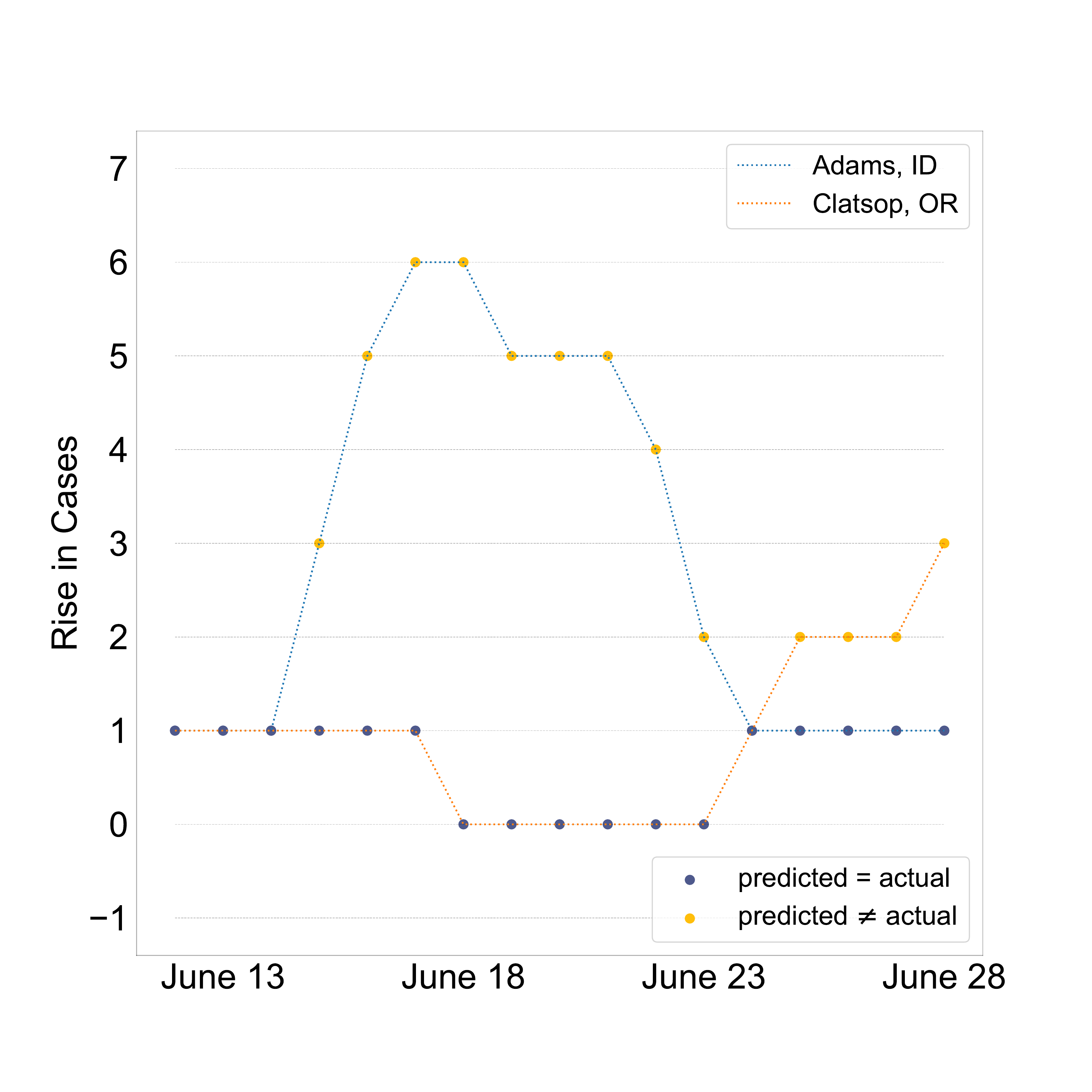}
\end{subfigure}
\caption{Example results for the model predictions over time for three types of counties: counties with high growth, counties with medium growth, and counties with low growth of cases.}
\label{fig:temporal}
\end{figure*}

\textbf{Feature Analysis.} For the model to be informative for policy making, it should be explainable. To this end, we adopted a simple and popular method to evaluate feature importance \cite{samek2017explainable}. We first evaluated the accuracy of the model on a small section of the training set. Next, we looped through every feature in every feature group, randomized its value, and reevaluated the model's accuracy on the same section of the training set. We assigned higher importance to those features which when randomized cause greater decrease in the model accuracy.

A similar analysis was performed to determine feature values at which time steps are the most important in determining the final output. To perform this analysis, we randomized all features on a given time step, and then assigned greater importance to those time steps which when randomized cause greater decrease in performance.

\textbf{Feature Interaction Analysis.} As described in section \ref{sec:model}, the DeepCOVIDNet model explicitly computes second-order interactions between input feature groups. Due to this explicit representation of second-order interactions, it is possible to identify pairs of features between which high amount of interaction is observed. To perform this analysis, we evaluated the network on a section of the training set and tracked the magnitude of activations in the vector representing the second-order interactions of input features. We concluded that the observed interaction among two feature groups is high when the average magnitude of their second-order interaction is also high. This implicit assumption that neurons are activated highly when they capture a pattern to which they are responding to is common in deep learning and is used in other interpretation techniques, such as activation maximization \cite{erhan2009visualizing}.

\section{Results}

\subsection{Predictive Performance}
In this section, we discuss the performance of the proposed model on the test set. As discussed above, the test set contains 17 days of data from June 12 through June 28. The average accuracy of the model on these 17 days is 63.7\% when using four output classes to categorize the growth in the number of new cases for each county (i.e. negligible increase, moderately low increase, moderately high increase, and significantly high increase).

Further, as shown in figure \ref{fig:acc}, the performance of the model, which is trained from April 5 through June 11, decreases as we evaluate on dates further from June 11. The same trend of the model performance decreasing over time can also be observed in figures \ref{fig:spatial} and \ref{fig:temporal}. This could be due to two possible reasons. One, the COVID-19 situation is highly dynamic and the behavior of people and the adaptive strategies they use change frequently. For example, mask use in the United States has increased over time, especially after the Centers for Disease Control and Prevention (CDC) advised it \cite{goldberg2020mask}. So if the model were trained on data before mask use became prevalent, it would learn trends that will not hold true after masks become more widespread. Two, the testing capacity of the United States continues to ramp up with time. As testing becomes more accessible, the trends a model may have learned earlier may not hold, as more people would be tested, resulting in more infected cases being found.

Therefore, during deployment, a good strategy is to let the model forecast only 7 days beyond the end of dates used in training, as new features are released about every 7 days. To project further into the future, the model must be retrained on the most recent training data available to ensure that the knowledge it learns holds for the dates it is asked to forecast on. It should be noted that research \cite{holmdahl2020wrong} has shown that predictive performance of many other forecasting models also declines with time. Therefore, since the model will only forecast 7 days into the future in practice, it is important to know its ``use case accuracy'' or ``7-day accuracy,'' which is the model's accuracy from June 12 through June 18 (7 days of test data). The model's "use case accuracy" is much higher at 70.8\%.

\begin{figure*}
\centering
\includegraphics[scale=0.32]{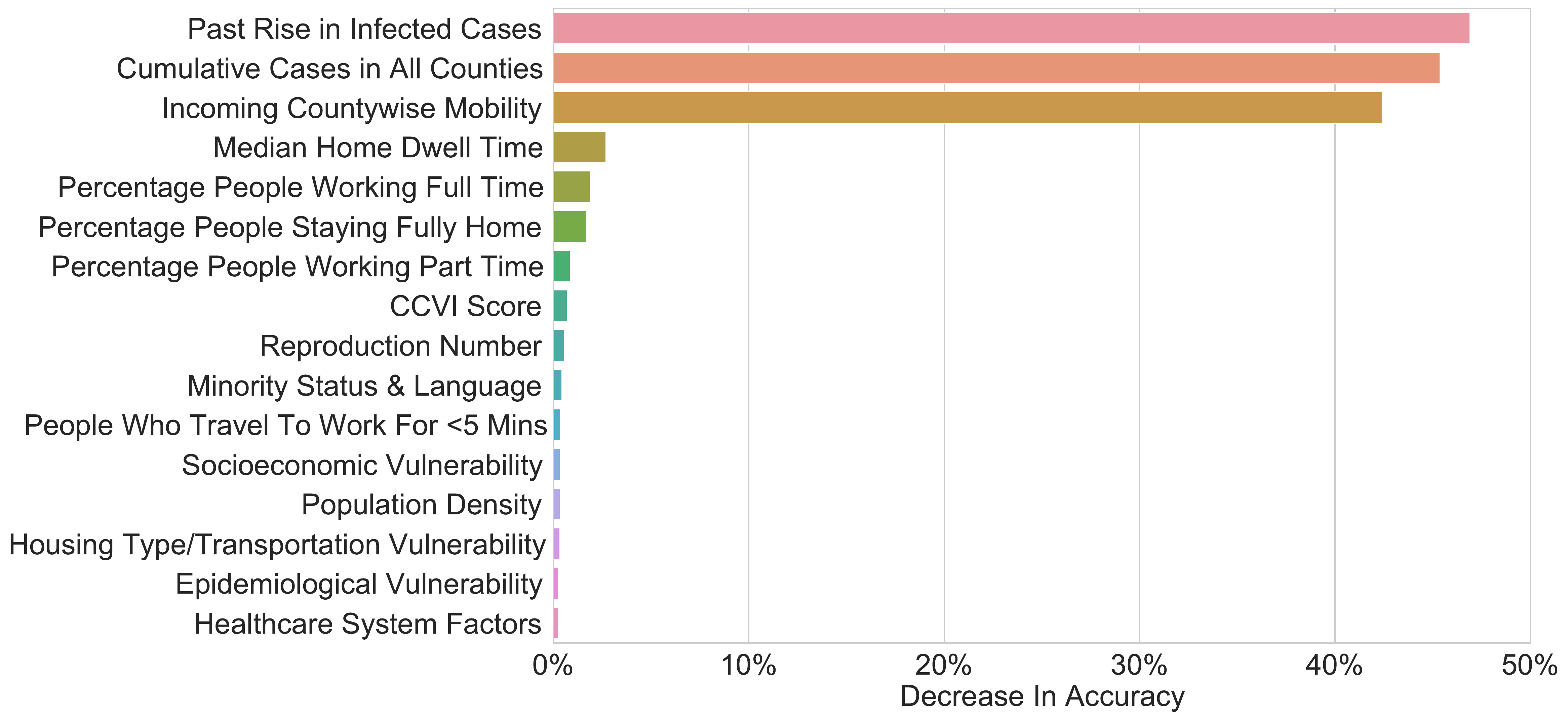}
\caption{Feature importance results (the greater the decrease in accuracy, the more important the feature).}
\label{fig:feature-analysis}
\end{figure*}

Figure \ref{fig:spatial} shows the predictive performance of the model in all U.S. counties on a series of dates. The figure shows that the majority of the predicted results are consistent with or one class away from the actual growth in the number of cases. The predicted growth for only a few counties are more than two or three classes away from the actual situation. This shows that the model generally performs well in predicting growth in spread of infection in a county. Even when the model's predictions are incorrect, they are usually just one category away from being correct. Further, as discussed above, the figure also shows that the accuracy of the model decreases for dates further away from those used in the training set.

Figure \ref{fig:temporal} shows the model predictions over time for different counties based on the growth of infected cases in them. We see the same general trend as in figure \ref{fig:spatial} that the model makes more inaccurate predictions as we go further in the future. 

\subsection{Feature Importance Evaluation}

\begin{figure}
\includegraphics[scale=0.53]{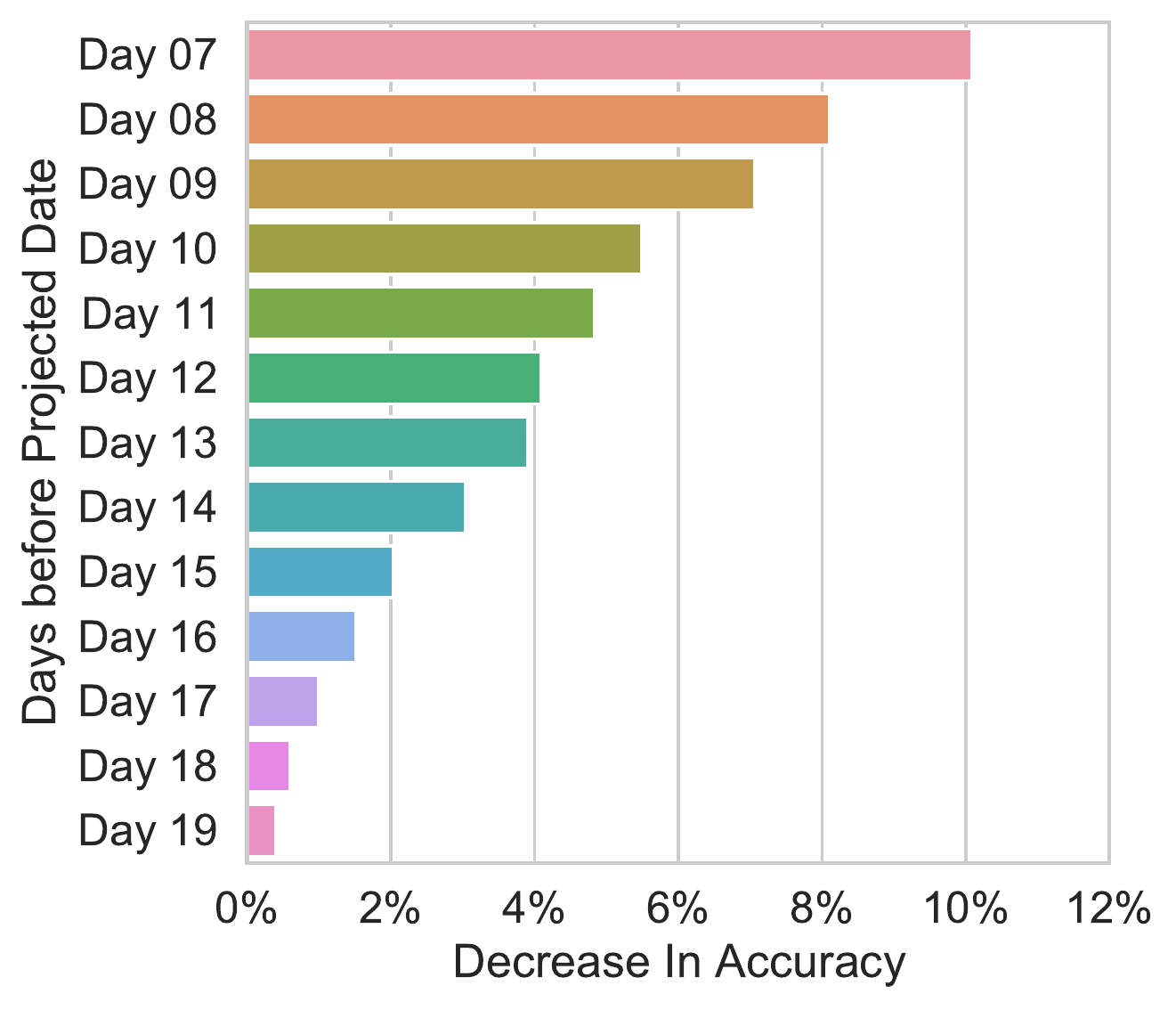}
\caption{Time step analysis. This analysis shows how important features of each of the past days are in predicting the final output. The results indicate that the feature values related to the most recent days are most important in predicting the growth of cases 7 days in the future. Refer \ref{ssec:ftr-analysis} for more details.} 
\label{fig:time-step-analysis}
\end{figure}

\label{ssec:ftr-analysis}
In the next step, we evaluated the importance of different county features in terms of their contribution to the prediction of the number of new cases in each county. Figure \ref{fig:feature-analysis} shows some of the most salient features identified by the model using the method described in section \ref{sec:impl}. As shown in the figure \ref{fig:feature-analysis}, the most important feature is the past increase in infected cases in the current county. This implies that the past trend of the growth of infection in a county is a strong determinant of future growth. Moreover, the second and third most important features are the cumulative infected cases in all counties and the number of people from other counties visiting the county under study. As explained in appendix \ref{sup:ftr}, these are both cross-county time-dependent features. Together, these features could imply that the phenomenon of people traveling from more infected counties to other counties is associated with the growth of infected cases in the destination county. In other words, the results show that inter-county travel risk could be high for the destination county, consistent with previous studies about the risks of travel \cite{rodriguez2020going} \cite{lai2020assessing}. Other important features, such as the median home dwell time, percentage of people working full- or part-time, percentage of people staying home, and number of people traveling to work for less than 5 minutes all show that the model has learned the importance of staying home and social distancing in general, which is again consistent with the reported findings in previous studies \cite{ferguson2020report}. In addition, the model also shows that various manually engineered features are effective and indicative of the growth of infected cases. For example, COVID-19 Vulnerability Index (CCVI) score, socioeconomic vulnerability, epidemiological vulnerability, and healthcare system factors are engineered features designed by the Surgo Foundation \cite{ccvi} to identify communities that are at a greater risk of infection, and our results show that all these features are important. Note that other features, such as estimated reproduction number (equal to the number of other infections caused by one infected person) and population density, also make the list of important features.

Figure \ref{fig:time-step-analysis} shows the results of the time-step analysis so we know features of which of the past days are most influential in predicting the final output. As discussed earlier, all time-dependent and cross-county time-dependent features use 13 days of past data. Since the projection interval of the model is 7 days, the most recent day used for input features is 7 days before the prediction date (also called projected date). As shown by figure \ref{fig:time-step-analysis}, the most recent days are the most important in predicting the growth of cases.

\subsection{Feature Interactions}
\label{ssec:ftr-int}

\begin{figure}
\centering
\includegraphics[scale=0.4]{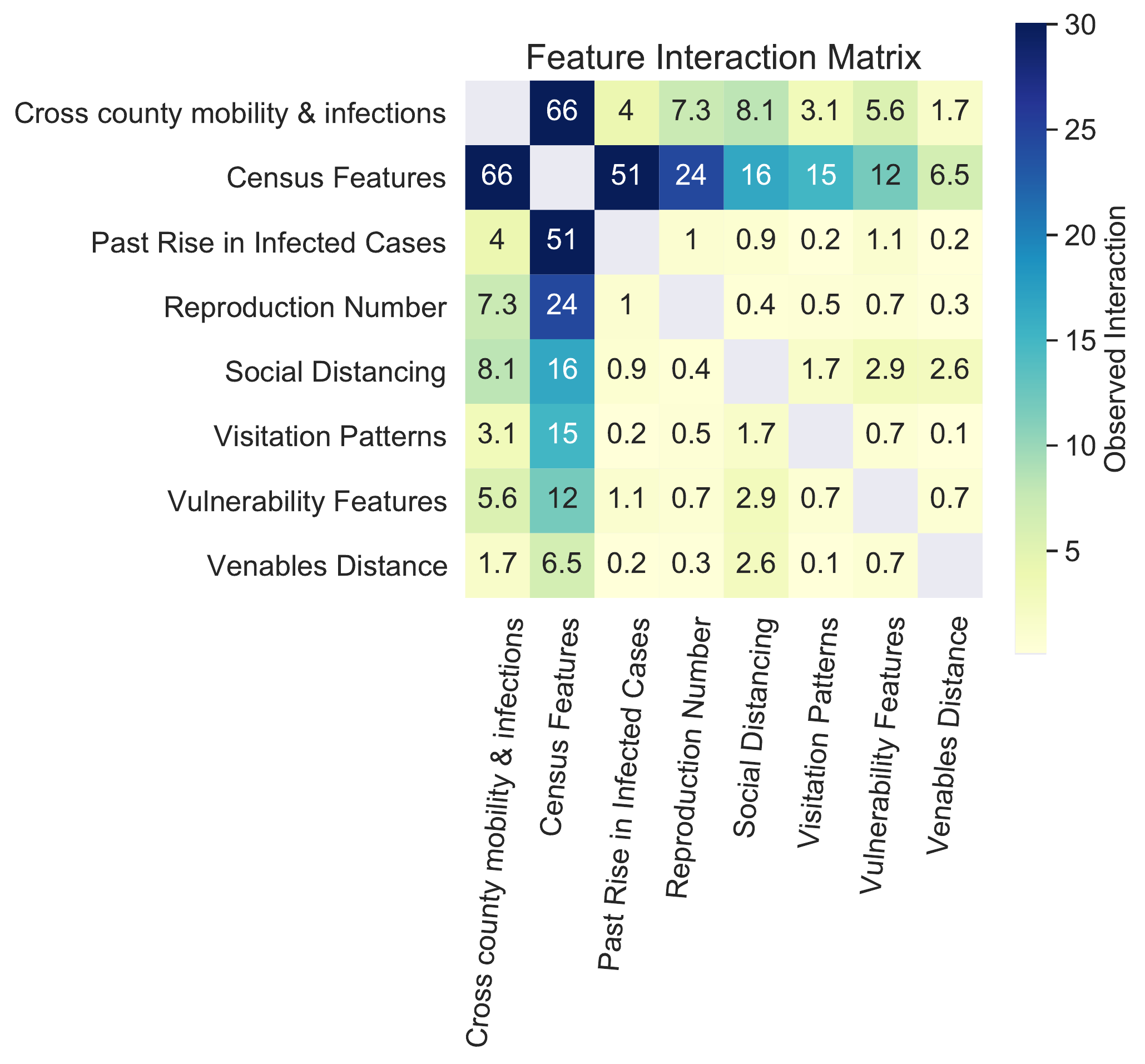}
\caption{The observed second-order interactions between input feature groups. Higher value in a cell represents higher amount of interaction among the corresponding feature groups. Refer to \ref{ssec:ftr-int} for more details.} 
\label{fig:ftr-int}
\end{figure}

Figure \ref{fig:ftr-int} shows second-order interactions between all feature groups computed by the method described in section \ref{sec:impl}. The values in the cells represent the mean activations of the second-order interactions of corresponding feature groups. As discussed in section \ref{sec:impl}, higher values represent higher observed interactions. An important observation from the figure is that almost all feature groups have notable interaction with census features. This result indicates that all other county-level features must be interpreted in the context of the population attributes of that county. For example, two counties with different population attributes may require varying amounts of adherence to social distancing orders to produce the same dampening effect in the growth of disease spread. Note also that figure \ref{fig:feature-analysis}, which shows important features identified by the model, does not include many census features, indicating that census features alone do not contribute much to prediction of the growth in future cases.

Also, the results show that many feature groups have relatively high interaction with cross-county mobility and infections. For example, a relatively high interaction exists between cross-county mobility \& infections and social distancing metrics. This result indicates that travellers from other counties have a different kind of impact in destination counties which have higher adherence to social distancing than they do in destination counties with lower adherence levels. This analysis validates our initial hypothesis that interactions among the many input feature groups are likely to exist and aid in predicting the rise in future number of cases in counties.

\section{Discussion and Concluding Remarks}
In this paper, we propose a novel deep learning model to examine heterogeneous county-level features and to predict growth of infected cases in the future. The proposed method can extract embeddings from multivariate time series and multivariate spatial time series data in a novel way by utilizing both the temporal and spatial (if available) structure of the data. This process of extracting embeddings could be employed by other deep learning research to process similar type of data. Further, unlike existing models \cite{tomar2020prediction} \cite{perc2020forecasting} \cite{anastassopoulou2020data} \cite{huang2020multiple} \cite{chimmula2020time} \cite{punn2020covid}, the proposed model takes in a large number of input features and learns interactions between them. The application of the model has been demonstrated in predicting the growth in the number of new cases in U.S. counties during the COVID-19 pandemic. The model has acceptable performance in prediction and provides highly interpretable feature analysis results that can help policymakers cope with spread.

The proposed forecasting model can be effectively used for predictive pandemic surveillance by both governments and industries. For example, hospitals in each U.S. county can use the model to estimate the growth in the number of cases and determine their needs of future supplies and resources based on the projected number of infections. Similarly, schools and offices can make better plans for future weeks to be prepared in the best way. In addition, policymakers can use this model to deploy proactive measures instead of taking a reactive approach to dampen the spread of the virus and possibly save lives.

The findings of feature importance evaluations can further help researchers decide which features to include in their forecasting models. The results also open up new avenues for other researchers to do a detailed statistical analysis on how the most important features identified by our model exactly contribute to the growth of infected cases. Further, the proposed model can also be used to test the effectiveness of various hand engineered features in predicting the growth of the virus. For example, our results have shown that several manually engineered features like COVID-19 Vulnerability Index (CCVI) is effective in predicting the spread of the virus. Similarly, newly developed features created by other researchers can be added to our model and the feature analysis results can be relied upon to show relative importance of those features. Lastly, due to the high interpretability of our feature importance results, policymakers can design effective control strategies to prevent the growth of infection based on controlling the important factors identified by this study.

Similarly, the results of identifying pairs of features with high second-order interactions show that interactions among the many input features exist and are important to explicitly account for by future researchers in their forecasting models. Further, this analysis also opens up opportunities for data scientists and statisticians to explain more clearly the reasons for such interactions and their possible implications to epidemiologists studying the spread of COVID-19. 

In addition, we believe that data-driven models like the DeepCOVIDNet mark the advent of an era of extensively using deep learning for pandemic forecasting, which can complement existing epidemiology models. Already, there is an increasing interest in using recurrent networks to forecast the spread of COVID-19 \cite{ayyoubzadeh2020predicting} \cite{tian2020covid} \cite{yang2020modified}. To the best of our knowledge, the proposed model is among the earliest non-recurrent and non-conventional deep learning models used for pandemic prediction, which should encourage other researchers to develop more creative models. Finally, this work also presents one of the first attempts to make the results of a deep learning based COVID-19 forecasting model interpretable to policymakers and the public. 

Further improvement regarding the prediction accuracy and interpretability of the model are important future directions for this work. For example, although we explicitly model second order interactions between feature groups, we still lack a good way to interpret these interactions at the individual feature level. In other words, although we may be able to tell that there is high interaction between census features and social distancing features, we cannot tell which exact census features interact with which exact social distancing features. A strong understanding of which individual features interact will be of great value to epidemiology researchers and policy makers. Further, it is important to continue to add and engineer new features for the model and improve its predictive performance in future studies.

\section{Data and Code Availability}
The data that support this study are available from SafeGraph, Mapbox, and The New York Times GitHub repository containing information about the past number of cases. Restrictions apply to the availability of some of these data, which were used under license for the current study, and so are not publicly available. However, the full code used to implement the proposed DeepCOVIDNet model is available on Github: \url{https://github.com/urban-resilience-lab/covid-county-prediction}.

\section{Acknowledgment}
This material is based in part upon work supported by the National Science Foundation under Grant Number SES-2026814 (RAPID), the National Academies’ Gulf Research Program Early-Career Research Fellowship, the Amazon Web Services (AWS) Machine Learning Award, and Microsoft AI for Health COVID-19 Grant for cloud computing resources. We also would like to acknowledge the support from Mapbox \cite{mapbox} and Safegraph \cite{sgocd} \cite{sgp} \cite{sgsd} in providing data. Any opinions, findings, conclusions and/or recommendations expressed in this material are those of the authors and do not necessarily reflect the views of the National Science Foundation, Amazon Web Services and Microsoft.

{\small
\bibliographystyle{unsrt}
\bibliography{covid}
}

\section{Appendix: Feature Descriptions}
\label{sup:ftr}
In this section, we provide detailed descriptions of the feature groups used in the model. As shown in section \ref{sec:in-features}, the feature groups of the model are categorized based on four influencing factors. In this section, we provide the definition and detailed calculation of all feature groups based on their influencing factors.

\subsection{Population Attributes}
As mentioned in table \ref{tab:themes}, we use the following features to capture population attributes:

\begin{enumerate}
    \item \textbf{Census Features:} We use a subset of 2100 sociodemographic features compiled by SafeGraph \cite{sgocd}. These features include population level information about age/sex, race, ethnicity, commuting information, household/family type, school enrollment, language spoken at home, poverty status, income, employment status and occupation. This data is originally collected by the American Community Survey in 2016.
    \item \textbf{Vulnerability Features:} In this group, we add population density and other features built by the Surgo Foundation \cite{ccvi} to identify counties that are more vulnerable to the spread of the virus. Specifically, the foundation has created seven features, six of which assess vulnerability of a county across different areas and one assesses overall vulnerability to COVID-19 as described in the following. All features are bounded between 0 and 1, where 0 means least vulnerability and 1 means greatest vulnerability. It is important to note that these features are defined in a relative context, meaning that the value 1 is associated with the most vulnerable county. Due to this reason, the "feature values" actually represent vulnerability rankings of all counties. Each of the seven features shown below is developed by combining multiple other features as described below.
    \begin{enumerate}
        \item \textbf{Socioeconomic Vulnerability.} This feature is developed based on population below poverty line, unemployed population, per-capita income, and population without high school diploma and represents the socioeconomic vulnerability of a county.
        \item \textbf{Household Composition \& Disability.} This feature is developed based on the distribution of population older than 65, population younger than 17, population with a disability, and single parent households. It is a measure of vulnerability of households and the population at large.
        \item \textbf{Minority Status \& Language.} As the name suggests, this feature is developed based on minority population and number of people who speak ``less than well'' English.
        \item \textbf{Housing Type \& Transportation.} This feature is developed by using the estimate of mobile homes, households with more people than rooms, households with no available vehicle, housing with structures with 10 or more units, and persons living in institutionalized group quarters. This feature helps identify communities with poor housing or transportation situation.
        \item \textbf{Epidemiological Vulnerability} This feature identifies communities with greater risk of negative impact during disease epidemics. It is built by using 11 other features capturing the number of people with cardiovascular issues, respiratory conditions, weak immune systems, obesity, diabetes, and high vulnerability to influenza and pneumonia. The feature also considers population density of the county.
        \item \textbf{Healthcare System Factors.} This feature tries to measure the capacity, strength (measured by county spending on health and research quality) and preparedness of the health care system in a county using 8 different features. 
        \item \textbf{COVID-19 Vulnerability Index (CCVI).} Finally, the above 6 features are combined with equal weighting to create the COVID-19 vulnerability index, which is intended to identify communities ``with a limited ability to mitigate, treat, and delay the transmission of'' the virus.
    \end{enumerate}
\end{enumerate}

\subsection{Population Activities}
We use the mobility data collected by SafeGraph to capture population activities. The SafeGraph Patterns \cite{sgp} and SafeGraph Social Distancing Metrics \cite{sgsd} are adopted in our use case. The former includes information related to the number of visits to certain points of interest, and the latter provides data to show adherence to stay-at-home guidelines. 
\begin{enumerate}
    \item \textbf{Visitation Patterns}: SafeGraph Patterns data has information about the number of people that visit different types of points of interest (i.e. grocery stores, health facilities, etc) on a given day. We consider visits to the following types of places in our model: amusement parks and arcades, colleges/universities/professional schools, living facilities for elderly, department stores, hospitals, merchandise stores/supercenters, grocery stores, and restaurants/eating places. As discussed in \ref{ssec:feature-rationale}, visits to grocery stores, restaurants/other eating places, living facilities for elderly, and hospitals could be particularly helpful in forecasting the growth of cases since these places could facilitate virus transmission \cite{benzell2020rationing} \cite{lai2020covid}  \cite{bahl2020airborne} \cite{morawska2020can}.
    \item \textbf{Social Distancing Metrics}: We use SafeGraph social distancing metrics to compute/extract the percentage of people staying home, percentage of people working full- or part-time, median distance travelled from home and the median amount of time spent at home for each county and each day. As discussed in \ref{ssec:feature-rationale}, adherence to stay-at-home guidelines and the number of people working outside are both important in determining the spread of the virus \cite{gao2020mobile} \cite{cohen2020countries} \cite{sen2020social} \cite{beland2020short}.
    \item \textbf{Venables Distance}: Venables distance captures the concentration of population activities in a county. Venables distance, $D_V(t)$, is defined formally as \cite{louail2014mobile}:
    
    \begin{align}
        D_V(t) = \frac{\sum_{i < j} s_i(t) s_j(t) d_{ij}}{\sum_{i < j} s_i(t) s_j(t)}
    \end{align}
    where $s_i, s_j$ represents the population activity intensity in cell $i$ and $j$ respectively and $d_{ij}$ represents the distance between the two cells. In our analysis, we define a cell to be a 4 km\textsuperscript{2} area. We compute $D_V$ for each day and county by basically using the daily average values for $s_i(t)$ and $s_j(t)$. Louail et al. \cite{louail2014mobile} describe more details about this equation.
    We use Mapbox digital trace telmetry data, which basically uses aggregated cell phone data to estimate population activity in small spatial regions of each county, to extract values of $s_i(t)$ and $s_j(t)$ for cells.
    
\end{enumerate}

\subsection{Mobility}
To capture mobility across regions, we employed inter-county travel data from SafeGraph as described below.

\begin{enumerate}
    \item \textbf{Cross County Mobility \& Infections}: The SafeGraph data \cite{sgp} include the census blocks where the travelers visiting a certain point of interest come from. With this information, we can estimate the total number of travelers from one county to another. We further augment the mobility data by also adding the cumulative number of infections in source counties, so the model can realize that visitors from a more infected county could be more dangerous than visitors from a less infected county.
\end{enumerate}

\subsection{Disease Spread Attributes}

\begin{enumerate}
    \item \textbf{Past Rise in Infected Cases}: We consider the weekly rise in the number of confirmed cases in past days as a feature to represent the past pandemic situation in a county. This data is obtained from the New York Times GitHub repository \cite{nytimesdata}. Since the proposed model predicts the growth of cases a week in the future, the weekly rise in cases for the past few days as an input feature can inform the model about the recent trend of the variable that it has to predict.
    \item \textbf{Reproduction Number}: The basic reproduction number is also used, which is an estimate of how fast the number of cases are expected to increase by. Formally, it is the estimated number of other cases caused by one infected person. Fan et al. \cite{fan2020effects} describe the exact method of estimating reproduction number by using a simple epidemic model. They assume that an infected person infects another $R_0$ people after time $\tau$. Therefore, if the number of infected people at time step 0 are $i(0)$, then at time step $t$, the number will grow to $i(0)R_0^{t/\tau}$. Simple algebraic manipulations show that:
    \begin{align}
        R_0 = e^{\frac{(\ln{i(t)} - \ln{i(0)}) \cdot \tau}{t}}
    \end{align}

    Fan et al. \cite{fan2020effects} set $\tau$ to 5.1 days and provide justifications for this choice in their work. As table \ref{tab:in-features} shows, we use reproduction number as a time dependent feature in the model. Therefore, for this analysis, we estimate the daily reproduction number based on 10 days of past data. In other words, to estimate $R_0$ at day $d$, we use the following formula:
    \begin{align}
        R_0(d) = e^{\frac{(\ln{i(d)} - \ln{i(d - 10)}) \cdot \tau}{d}}
    \end{align}

\end{enumerate}

\end{document}